\documentclass[aip,jcp,reprint]{revtex4-1}

\newcommand{\me}{\mathrm{e}}

\newcommand{\dif}{\mathrm{d}}

\newcommand{\lab}{\left<}
\newcommand{\rab}{\right>}

\newcommand{\exmu}{\mu^{(\mathrm{ex})}}  
\newcommand{\ep}{\varepsilon}        

\usepackage{graphicx}

\usepackage{amsmath}

\usepackage{amsfonts}

\usepackage{dcolumn}

\usepackage{bm}

\begin{document}

\title[Interfaces of Propylene Carbonate]{Interfaces of Propylene Carbonate}

\author{Xinli You}\email{xyou@tulane.edu}
\author{Mangesh I. Chaudhari}\email{chaudhari.84@gmail.com}
\author{Lawrence R. Pratt}\email{lpratt@tulane.edu}
\author{Noshir Pesika}\email{npesika@tulane.edu}
\affiliation{Department of Chemical and Biomolecular Engineering, Tulane
University, New Orleans, LA 70118}

\author{Kalika M. Aritakula}\email{karitaku@uno.edu}
\author{Steven W. Rick}\email{srick@uno.edu}
\affiliation{Department of Chemistry, 
University of New Orleans, New Orleans, LA 70148}

             
\begin{abstract} 

Propylene carbonate (PC) wets graphite with a contact angle of
31$^\circ$ at ambient conditions.   Molecular dynamics simulations agree
with this contact angle after 40\%  reduction of the strength of
graphite-C atom Lennard-Jones interactions with the solvent, relative to
the models used initially.   A simulated nano-scale PC droplet on
graphite displays a pronounced layering tendency and an \emph{Aztex
pyramid} structure for the droplet.  Extrapolation of the computed
tensions of PC liquid-vapor interface estimates the critical temperature
of PC accurately to about 3\%.  PC molecules lie flat on the PC
liquid-vapor surface, and tend to project the propyl carbon toward the
vapor phase. For close PC neighbors in liquid PC, an important packing
motif  stacks carbonate planes with the outer oxygen of one molecule
snuggled into the positively charged propyl end of another molecule so
that neighboring molecule dipole moments are approximately antiparallel.
The calculated thermal expansion coefficient and the dielectric
constants for liquid PC agree well with experiment.  The distribution of
PC molecule binding energies is closely Gaussian. Evaluation of the
density of the coexisting vapor then permits estimation of the packing 
contribution to the PC chemical potential, and that contribution is
about 2/3rds of the magnitude of the contributions due to attractive
interactions, with opposite sign. 

\end{abstract}

\maketitle

\section{Introduction}

\begin{figure}
         \begin{center}
             \includegraphics[width=3.2in]{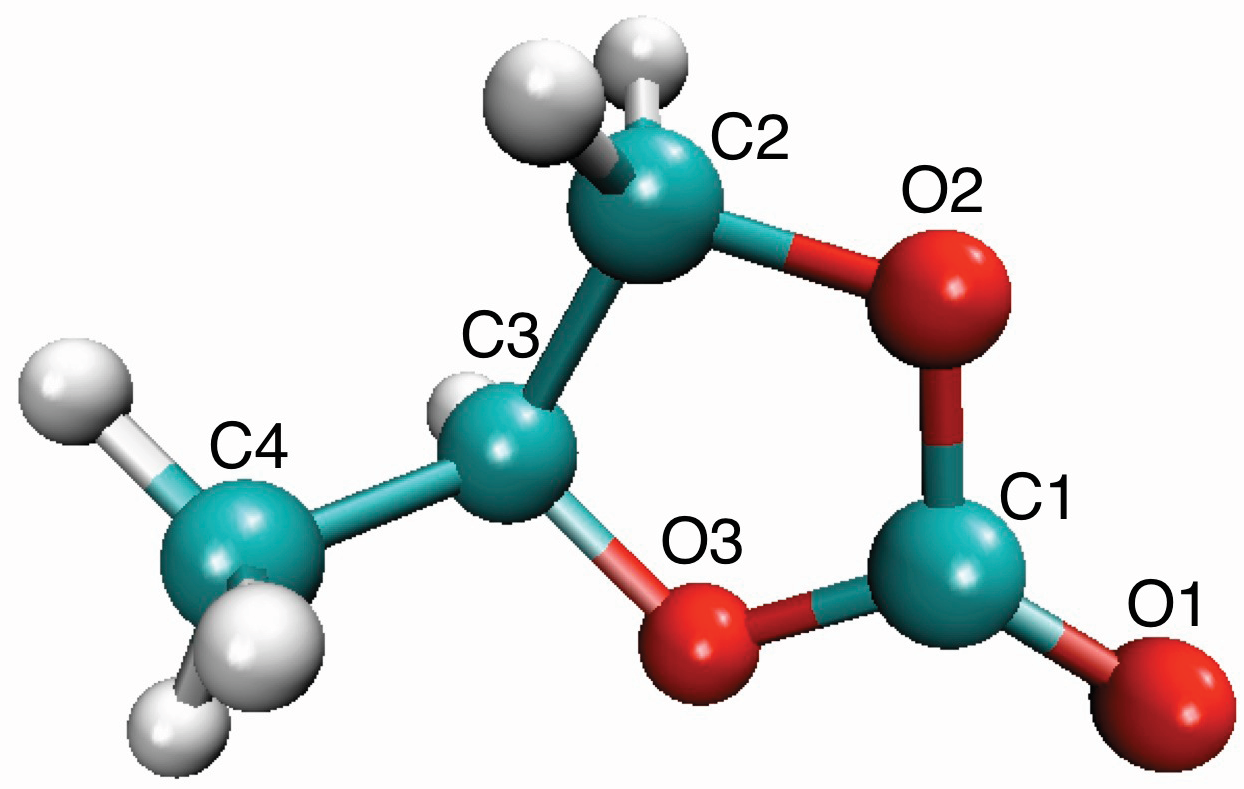}
         \end{center}
         \caption{Propylene carbonate (PC) enantiomer with atom
         labelling used in this paper.
         \label{fig:pc_labelled}}
\end{figure}

Unique properties of nanomaterials, and specifically of electrochemical
double-layer capacitors (EDLC) based on carbon nanotube (CNT)
forests,\cite{Dillon:2010bb} arise from their large surface areas.
Accurate descriptions of interfaces is an important challenge for
modeling. Here we considered propylene carbonate (PC:
FIG.~\ref{fig:pc_labelled})  as a solvent for electrochemical double-layer
capacitors, reporting experimental and molecular simulation
results on PC interfaces. This validation is preparatory to direct 
simulation of dynamical filling and performance of CNT-based EDLCs.

Differential filling of a charged nanotube forest from a bulk
electrolyte solution is expected to be  sensitive to the balance of
attractive intermolecular interactions. A possible concern for the first
simulations of CNT-based EDLCs was that models of the PC solvent had not
been parameterized to describe contact with carbon
electrodes.\cite{Yang:2009ty,Yang:2010hd} An initial molecular dynamics
calculation showed complete spreading of the modelled liquid PC on
graphite,\cite{lasigma2011} in disagreement with experimental
observation of a PC droplet on graphite reported below. That
experience\cite{lasigma2012} suggested the advantage of studying PC
interfaces in validation of simulation models for electrochemical
applications.  The present work adjusts the Lennard-Jones interactions
with the graphitic carbon to agree with the observed contact angle.

Propylene carbonate is a non-reactive, low-toxicity, aprotic, highly
polar dielectric solvent widely used in nonaqueous electrochemical
systems. Simulation of interfaces of non-aqueous solvents for
electrolytes, acetonitrile\cite{Paul:2005cn,Hu:2010iy} and
propionitrile,\cite{Liu:2012bl} have been previously carried-out.
Helpful molecular
simulations\cite{Soetens:2001ve,Silva:2007gg,Yang:2010hd} of the uniform
liquid propylene carbonate solvent are also available. A variety of
molecular dynamics results are discussed below, and technical details of
the methods are collected in the Appendix (Sec~\ref{Methods}).

\section{Results}\label{sec:results}

\begin{figure}
         \begin{center}
             \includegraphics[width=3.0in]{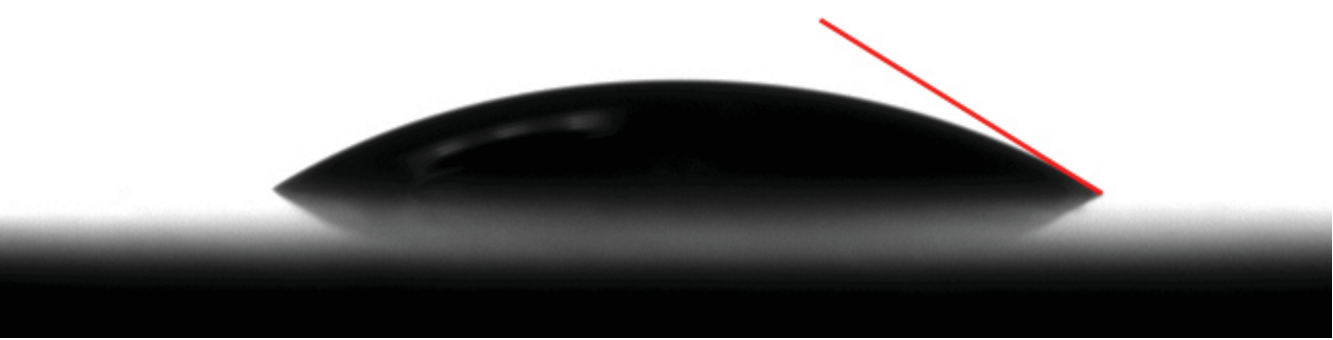}
         \end{center}
         \caption{PC droplet on graphite. Contact angle measurements of
         5 $\mu$l propylene carbonate drops on a pyrolytic graphite
         disc (Grade: PG-SN, Graphitestore.com) were performed under
         ambient conditions using a Rame-Hart contact angle goniometer. 
         The graphite disc was sheared against a paper wipe (Kimwipe) to
         reveal a fresh surface.  The surface was then rinsed with water
         and allowed to air dry.  A static contact angle of
         31.4$\pm$1.6$^\circ$ was obtained from 3 trials.
         \label{fig:contactAngle}}
\end{figure}

\subsection{Droplet-on-graphite contact angle}

\begin{figure}
         \begin{center}
             \includegraphics[width=3.2in]{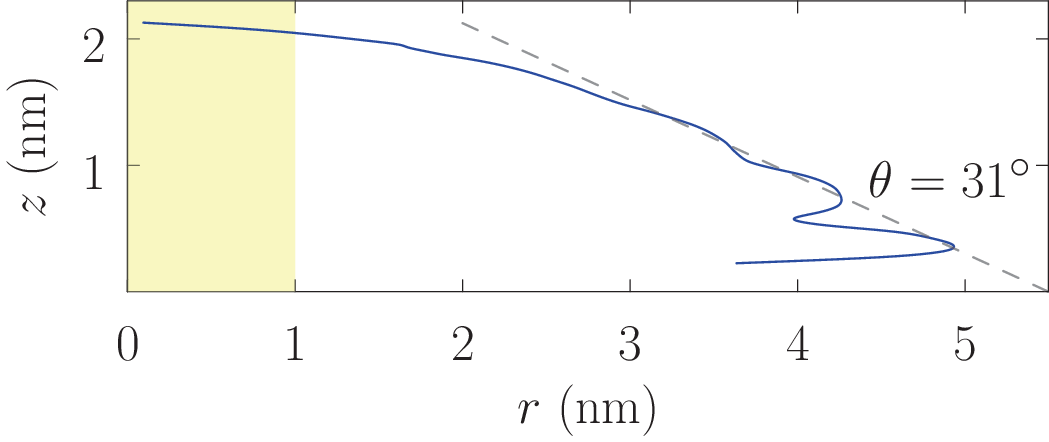}
         \end{center}
         \caption{Droplet silhouette, suggesting an Aztex
         pyramid.\cite{Heslot:1989tl,PGG} The indicated contact angle
         is found by fitting a straight line to the region
         $3.0~\mathrm{nm} < r < 4.2~\mathrm{nm}$. The shaded region
         identifies the cylinder used to investigate the layering of the
         mean density in FIG.~\ref{fig:layering}. \label{fig:aztexPyramid}}
\end{figure}

The observed contact angle (FIG.~\ref{fig:contactAngle}) is acute. This
indicates good wetting and favorable PC:graph\-ite interactions, though
not the complete spreading that was obtained\cite{lasigma2011} from
simulation with initial models.\cite{Yang:2009ty,Yang:2010hd}
Trial-and-error (Sec.~\ref{ref:dispersion})  found a simulation contact angle of about
31$^\circ$ with a 40\% lowering of the strength of Lennard-Jones
interactions associated with the C atoms of the graphite substrate
(FIG.~\ref{fig:aztexPyramid}).

Simulations of water on graphite surfaces show that changing the number
of water molecules from 1000 to over 17,000
has only a small effect on the contact angle.\cite{Werder:2003ve}
A droplet size dependences of such results are often ascribed to a
line-tension effect. The fine structure observed
here in the three-phase contact region here would make definition and
further investigation of a triple line complicated.  Together with the
distinctly non-spherical molecular shape and packing pattern (studied
below), interesting structural transitions of PC molecules near
\emph{charged} electrodes are expected; those issues should be the subject
of future studies.

\begin{figure}
         \begin{center}
             \includegraphics[width=3.2in]{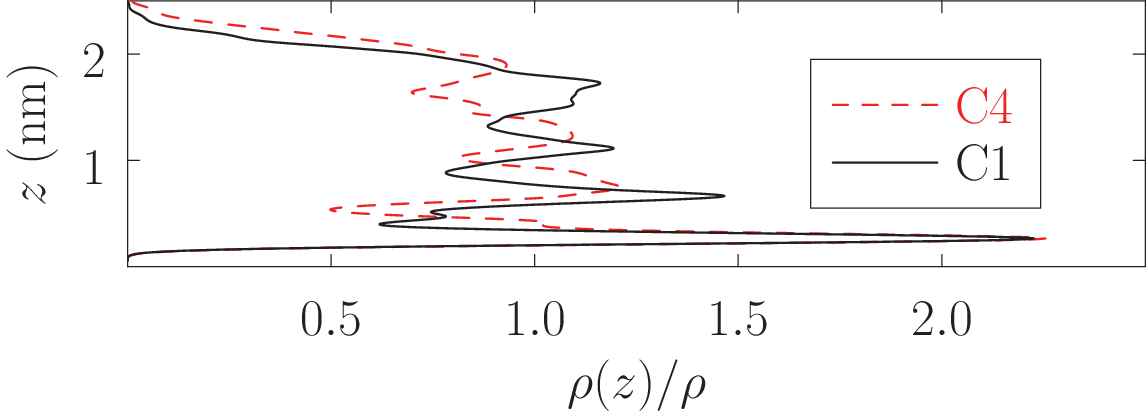}
         \end{center}
         \caption{The density, relative to the coexisting liquid density, in the
         central plug of material indicated by the shaded region of
         FIG.~\ref{fig:aztexPyramid}. Atom layers nearest to the
         graphite are concordant.   The density oscillations of C1 and
         C4 atoms conflict at larger distances from the graphite, and C4
         prevails  furthest from the graphite.
         \label{fig:layering}}
\end{figure}

The millimeter length scales of the experimental observation differ by
10$^6$ from nanometer length scales treated by the calculation. Pore
radii for ECDL capacitors are in the nanometer range. Therefore, the
nano-scale fine-structure of the molecular dynamics result
(FIG.~\ref{fig:layering}) should be relevant to the anticipated studies of
EDCLs.

\subsection{Liquid-vapor interfacial tensions}\label{sec:lvtension}

\begin{figure}
         \begin{center}
             \includegraphics[width=3.0in]{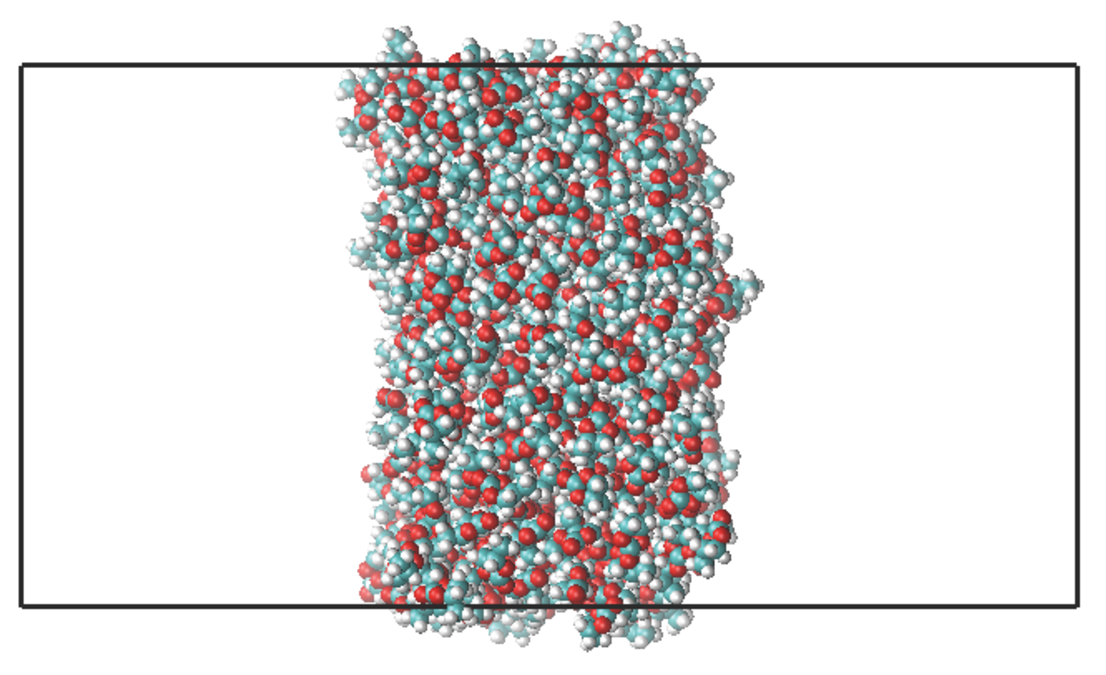}
             \includegraphics[width=3.0in]{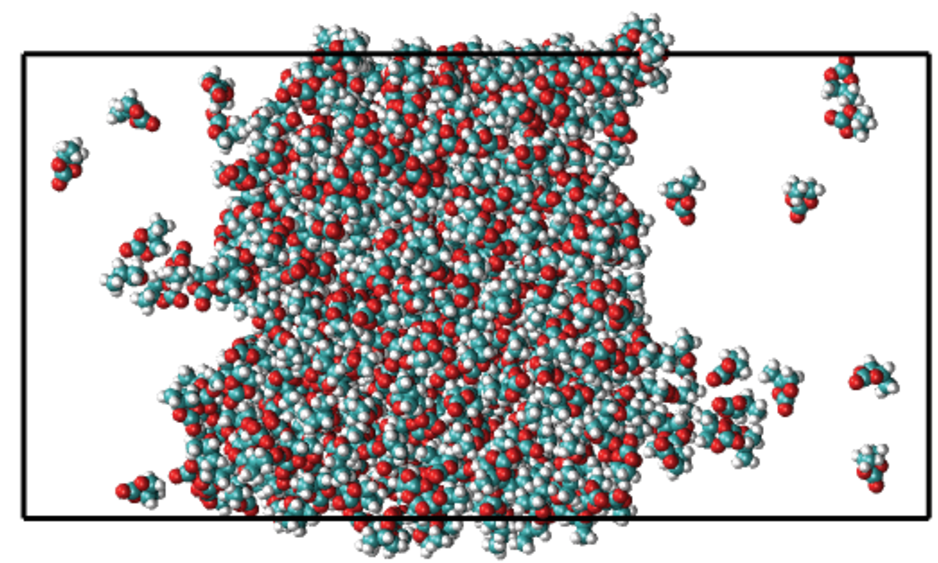}
         \end{center}
         \caption{Slab geometry, upper configuration from $T$=300K simulation, lower
         configuration from $T$=600K simulation.
         \label{fig:PurePC_3}}
\end{figure}

\begin{figure}
         \begin{center}
             \includegraphics[width=3.0in]{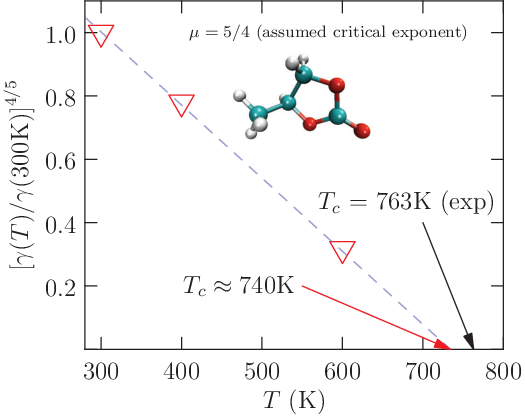}
         \end{center}
         \caption{Surface tensions  $\gamma =
         \frac{1}{2}\int \left\{p_{zz}(z) -  \frac{1}{2}\left\lbrack
         p_{xx}(z)+p_{yy}(z)\right\rbrack\right\}\dif z$. The
         experimental critical  temperature \cite{Wilson:2002dq} is   $T_{\mathrm{c}}$ = 763~K, 
         whereas the triple temperature is about 
         $T_{\mathrm{triple}}\approx 220$~K. \label{fig:gamma04}}
\end{figure}

Surface tensions were evaluated from simulations in slab
geometry (FIG.~\ref{fig:PurePC_3}) at several temperatures. The PC
liquid-vapor interfacial tensions (FIG.~\ref{fig:gamma04}) agree accurately
with experimental values (TABLE~\ref{tab:slabresults}) at moderate
temperatures. The simple formula $\gamma(T) \sim \gamma_0
\left(\frac{T-T_c}{T_c}\right)^{5/4},$ utilizing an approximate estimate
of the critical exponent,  extrapolates to $\gamma \rightarrow 0$ at
$T_c \approx $ 740K (FIG.~\ref{fig:gamma04}).

\begin{table*}
	\begin{center}
	\caption{Experimental surface tension at 20C is 41.1
	dyne/cm.\cite{Adamson1997} Vapor
	densities $\rho_{\mathrm{vap}}$ were obtained from FIG.~\ref{fig:FoldedProfileC1_TCompare},
	utilizing WHAM
	calculations at the lowest temperature.  \label{tab:slabresults}}
	\begin{tabular}{|c|c|c|c|c|c|c|}
	\hline
	   $T$(K) 
	& $\rho_{\mathrm{liq}}$ (nm$^{-3}$) 
	&  $\rho_{\mathrm{vap}}$ (nm$^{-3}$)  
	& $-kT\ln\left(\frac{\rho_{\mathrm{liq}}}{\rho_{\mathrm{vap}}}\right)$~(kcal/mol)
	&  $\rho_{\mathrm{vap}} kT$~(bar)
	&  $p$~(bar) \cite{Wilson:2002dq}
	&  $\gamma$~(dyne/cm) \\ 
\hline
300 & 7.11 & 1.6e-6 & -9.2 & 6.2e-5 & 7.6e-5 & 40.8   \\ 
\hline
400 & 6.46 & 4.8e-4 & -7.6 & 2.6e-2 & 2.8e-2 & 29.9 \\ 
\hline
600 & 4.93 & 7.0e-2 &-5.1 & 5.8 &  4.9 & 9.7 \\ 
\hline
\end{tabular}
\end{center}
\end{table*}


\subsection{Liquid-vapor interfacial structure}
\begin{figure}
         \begin{center}
             \includegraphics[width=3.0in]{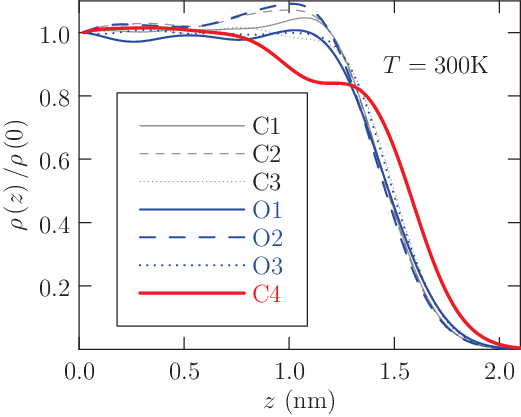}
         \end{center}
         \caption{The carbonate plane lies flat on the
         surface, projecting C4 from  the liquid toward the vapor phase.
         \label{fig:FoldedProfile_300K}}
\end{figure}

\begin{figure}
         \begin{center}
             \includegraphics[width=3.0in]{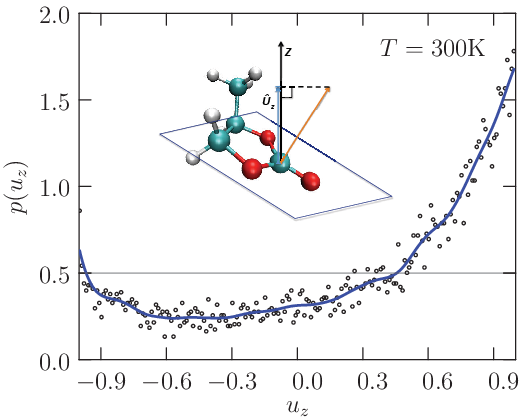}
         \end{center}
         \caption{For molecules with $z_{\mathrm{C1}}>$1.5~nm
         (FIG.~\ref{fig:FoldedProfile_300K}), the probability density 
         for projection of unit vector normal to
         the carbonate (-CO$_2$-) plane onto the $z$-axis, perpendicular
         to the interface. In this interfacial layer, the most probable
         orientation aligns the carbonate plane parallel with the plane
         of the interface, with the C4 methyl group extended toward the
         vapor phase.  $u_z > 0.5$ ($\theta < 60^\circ$) for about
         50\% of interfacial PC molecules.
         \label{fig:CarbonatePlane}}
\end{figure}

For the  lowest temperatures considered here, the interfacial profiles
of the atomic densities are distinctive (FIG.~\ref{fig:FoldedProfile_300K}): 
non-monotonic except for the density of the propyl carbon (C4).  The
coincidence of the positive lip in atom densities other than C4 with the
deficit in the C4 density suggests that the PC molecule lies
approximately flat on this interface while projecting C4 further toward
the vapor. Direct interrogation of the orientations of molecules with
$z_{\mathrm{C1}}>$1.5~nm (FIG.~\ref{fig:FoldedProfile_300K})  confirms this
view (FIG.~\ref{fig:CarbonatePlane}).

\begin{figure}
         \begin{center}
             \includegraphics[width=3.0in]{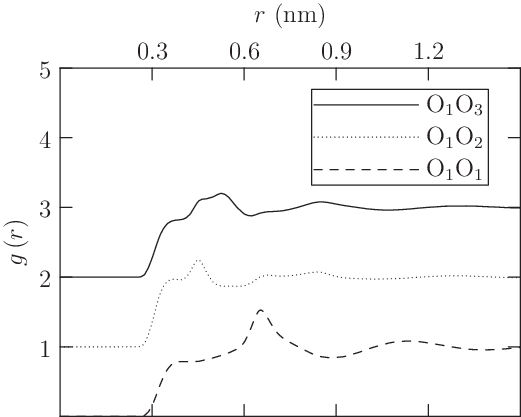}
             \includegraphics[width=3.0in]{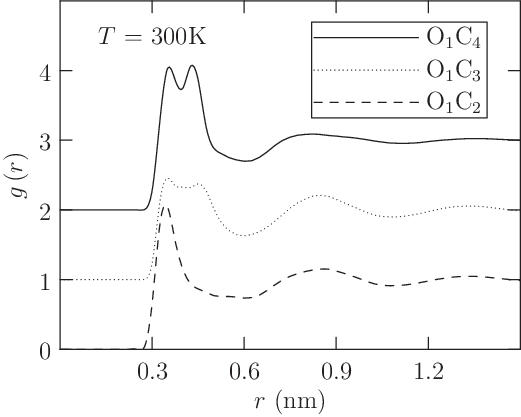}
         \end{center}
         \caption{Structured close-contacts involve negatively charged
         O1 with the opposite end of the PC molecule.
         \label{fig:grPC}}
\end{figure}

\begin{figure}
         \begin{center}
             \includegraphics[width=3.0in]{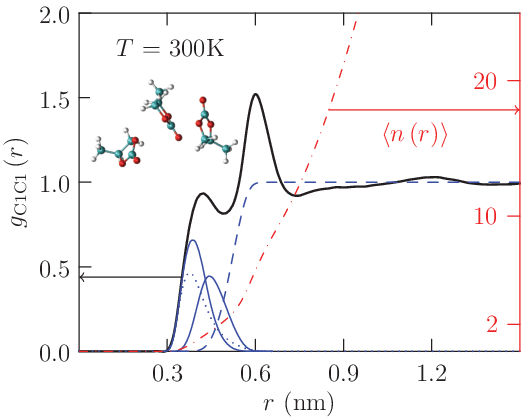}
         \end{center}
         \caption{Taking C1 as the center of the PC molecule, this
         characterizes the packing of centers, with molecular
         coordination numbers of 10 - 12. The solid blue curves are the
         radial distribution functions for the \emph{closest}, and
         \emph{2nd-closest} C1 neighbors of a C1 atom, peaked at 0.38~nm
         and 0.44~nm, respectively.  The dotted curve is the estimate of
         the radial distribution of the \emph{closest}  C1 neighbor of a
         C1 atom based on the Poisson approximation.\cite{Zhu:2011wn}
         The dashed blue curve is the probability, with median point of
         about 0.52~nm, that a central C1 atom has more than two (2) C1
         neighbors within that radius. The closest C1-C1 contacts are associated with
         stacking, with substantial disorder, of carbonate planes as
         suggested by the embedded molecular graphic.
         \label{fig:gnC1C1}}
\end{figure}

\subsection{Structure of the coexisting liquid}

The atom-atom intermolecular radial distribution functions
(FIGS.~\ref{fig:grPC} and \ref{fig:gnC1C1}) supplement the view that an
important packing motif  stacks carbonate planes of close PC neighbors
with the outer (O1) oxygen of one molecule snuggled into the positively
charged propyl end of another molecule so that neighboring molecule
dipole moments are approximately antiparallel. Such an arrangement is
similar to the known crystal structure of ethylene
carbonate,\cite{Brown:1954eo} and to the PC dimer structures established
by electronic structure calculations.\cite{Silva:2007gg}

On the basis of the C1C1 joint distribution (FIG.~\ref{fig:gnC1C1}), the
nearest two neighbors of a C1 are physically distinct, imperfectly
stacked on top and bottom of the carbonate plane.  The \emph{closest} C1
neighbor of a C1 atom is most probably at a radial displacement of
0.38~nm (FIG.~\ref{fig:gnC1C1}).  The observed distribution of the
\emph{closest} C1 neighbor of a C1 atom (FIG.~\ref{fig:gnC1C1}) is more
strongly peaked than the simple Poisson-based estimate.\cite{Zhu:2011wn}

\begin{figure}
         \begin{center}
             \includegraphics[width=3.0in]{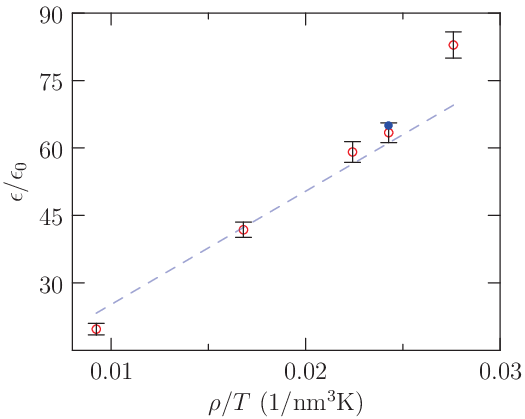}
         \end{center}
         \caption{Dielectric constant of model propylene carbonate,
         evaluated following standard simulation
         methods.\cite{Yang:2010hd,Wu:2006bg} The calculation treated
         600 PC molecules under periodic boundary conditions. Resulting
         values were averaged from the 40~ns production trajectories at
         constant pressure of 1~atm.  The
         solid dot is the measurement of C{\^{o}}te, \emph{et
         al.}\cite{Cote:96} at 25C.   Numerical values are given in
         TABLE~\ref{t:II}. \label{fig:eoT}}
\end{figure}

The dielectric constant of uniform PC liquid implied by these
simulations (FIG.~\ref{fig:eoT}) agrees satisfactorily with experiment.

\subsection{Balance of excluded volume and attractive interactions}

Thermodynamic energies for liquid PC may be analyzed on the basis of
quasi-chemical
theory\cite{Asthagiri:2007bl,Asthagiri:2010tj,Chempath:2009ws,%
Chempath:2009ut,Rogers2012} in which the paramount goal is molecular-scale
physical clarity from thermodynamic characteristics. The quasi-chemical
approach focuses on binding energies, $\varepsilon$ = $U(N) - U(N-1) -
U(1)$, of individual molecules and introduces a conditioning 
based on definition of an indicator function,\cite{ngvk} $\chi$, so that
$\chi=1$ defines a logical condition or constraint that physically
simplifies that statistical thermodynamic problem.\cite{Rogers2012}

For example, if close neighbors set particularly important or
complicated or \emph{chemical} interactions, then the logical   $\chi=1$
can indicate the absence of neighbors in a specified local region.   
Indeed, for  numerous applications to liquid
water,\cite{shah:144508,Chempath:2009ws,Chempath:2009ut,Rogers2012,%
Weber:2011hd,Weber:2010tg,Weber:2010bq}  $\chi=1$ indicates that
there are no O-atoms of bath molecules within a specified radius from
the O-atom of a distinguished water molecule. Then $\chi=0$ if any
solvent molecule \emph{is} present in that inner-shell.  For the 
application to CF$_4$(aq)\cite{Asthagiri:2007bl}, $\chi=0$ if any
water molecule is in defined van der Waals contact with the
polyatomic CF$_4$ solute.

With such a indicator function $\chi$
specified, the partial molar Gibbs free energy in 
\emph{excess} over the ideal contribution, $\exmu = \mu - \mu^{\mathrm{ideal}}$,  can
be expressed as
\begin{multline} 
\beta\exmu    =   -  \ln  \lab\lab \chi \rab\rab_{0} \\
+ \ln\int \me^{ \beta \ep }P(\ep | \chi =1) \dif\ep \\
+ \ln  \lab \chi \rab     ~,  
\label{eqn:fnrg1}
\end{multline}
with $\beta^{-1} = k_{\mathrm{B}}T$.   A principal virtue of this
formulation\cite{Rogers2012} is that is
subsumes a van der Waals picture of the solvation without requiring that
intermolecular interactions of different types be expressed is some
specific format.  We will use this formulation here to distinguish
packing contributions from longer-ranged interactions that are
attractive on balance. The notation $\lab \ldots \rab$ indicates the
usual average over the thermal motion of the system. The notation
$\lab\lab \ldots \rab\rab_{0}$ indicates the average over the thermal
motion of the system \emph{together} with an additional molecule with no
interaction between them, thus the subscript 0.   $\lab\lab \chi
\rab\rab_{0} $ is then the probability that the defined inner-shell is
empty in the case that interactions between the distinguished molecule
and the solution are absent.  That contribution thus gauges the free
energy cost of finding space for positioning the additional PC molecule
in the liquid.

More broadly, these formalities follow from the
identity\cite{Chempath:2009ws,Weber:2011hd} \begin{eqnarray}
\frac{\lab\me^{\beta\varepsilon}\chi\rab}{\lab\chi\rab} =
\me^{\beta\mu^{\mathrm{(ex)}}}\frac{\lab\lab
\chi\rab\rab_0}{\lab\chi\rab}~, 
\label{eq:broad}
\end{eqnarray} which is a disguised
expression of the \emph{rule of averages.}\cite{Paulaitis:2002vl,BPP}
This broader observation Eq.~\eqref{eq:broad}, might permit design of weight functions $\chi$ for
improved numerical performance.\cite{Chempath:2009ws}

\begin{figure}
         \begin{center}
             \includegraphics[width=3.0in]{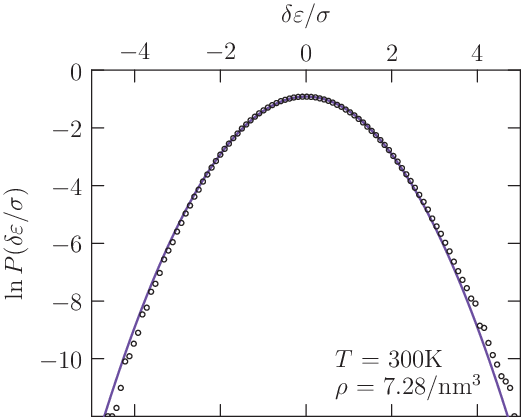}
             \includegraphics[width=3.0in]{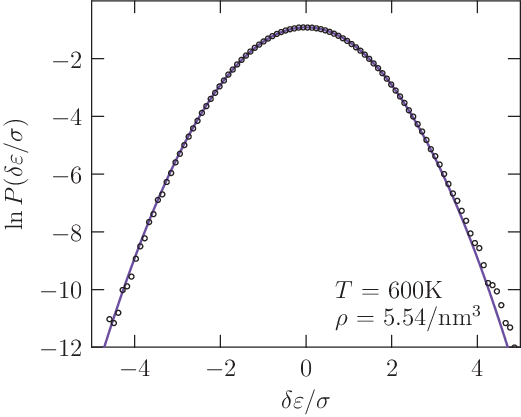}
         \end{center}
         \caption{Binding energies, centered ($\delta \varepsilon =
         \varepsilon - \left\langle \varepsilon \right\rangle$) and
         scaled  ($\sigma^2 = \left\langle  \delta
         \varepsilon^2\right\rangle$),   are approximately normally
         distributed. The slight super-gaussian behavior on the crucial
         right-side of these plots is the signature of repulsive
         intermolecular interactions. \label{fig:BE}}
\end{figure}

$P(\ep|\chi=1)$ that appears in Eq.~\eqref{eqn:fnrg1} is the probability
distribution of the binding energies, and is conditional on an empty
inner-shell.  That conditioning can make $P(\ep|\chi=1)$ simple enough
that a Gaussian (or normal) approximation
suffices.\cite{shah:144508,Chempath:2010dq}  The observed behavior of
$P(\ep)$ (FIG.~\ref{fig:BE}) already suggests that possibility.
Accepting that Gaussian approximation
\begin{multline}
\beta\exmu \approx -  \ln \lab\lab \chi \rab\rab_{0} \\
 + \beta\lab \varepsilon | \chi=1 \rab 
      	+ \beta^2\lab \delta \varepsilon^2 | \chi=1 \rab/2 \\
      	+ \ln \lab \chi \rab    ~.  
\label{eqn:fnrg2}
\end{multline}

For conceptual clarity,  let us discuss how definition of $\chi$ might
be approached.  $\chi$=1 corresponds to the absence of van der Waals
contact of the distinguished PC molecule with the solution.  	 We
might chose to define $\chi$ by assigning van~der~Waals radii for all
atom types.  The results of FIGS.~9 and 10, characterizing close
atom-atom pair distances, and particularly distance ordered
contributions of FIG.~10 would be directly relevant for that. We would
adjust those radii assignments to realistically large values, while
targeting values of $\lab \chi \rab$  not too different from one.

With this conceptual background, Eq.~\eqref{eqn:fnrg2} becomes
\begin{eqnarray}
\beta\exmu \approx -  \ln \lab\lab \chi \rab\rab_{0}
 + \beta\lab \varepsilon\rab 
      	+ \beta^2\lab \delta \varepsilon^2\rab/2   ~.  
\label{eqn:fnrg3}
\end{eqnarray}
This result is interesting for several reasons.  Though it
is suggestive of a van der Waals treatment, it is not limited to a
first-order mean-field contribution, and additionally the
assessment of attractive interaction is made on the basis of
observations for the case that those interactions are actually
operating. Nevertheless, the \emph{packing} contribution $-\ln \lab\lab
\chi \rab\rab_{0}$ is typically difficult to calculate directly.

\begin{figure}
         \begin{center}
             \includegraphics[width=3.0in]{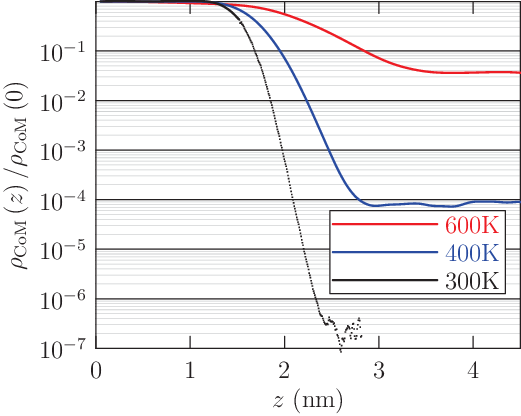}
         \end{center}
         \caption{The dotted results were obtained by the WHAM method
         in order to estimate of the vapor density  for the $T$=300K case
         where that value is not satisfactorily obtained from
         observation of the physical simulation.
         \label{fig:FoldedProfileC1_TCompare}}
\end{figure}

To quantitatively characterize the effort to obtain that
packing contribution, here we focus on getting an estimate on the basis
of the present methods.  With observation of the density of the vapor
phase (FIG.~\ref{fig:FoldedProfileC1_TCompare}), we can get the desired
free energy on the saturation curve\cite{BPP}
\begin{eqnarray}
\beta \exmu_{\mathrm{liq}}  = \beta\exmu_{\mathrm{vap}} -
\ln\left(\frac{\rho_{\mathrm{liq}}}{\rho_{\mathrm{vap}}}\right)~.
\label{eq:coex}
\end{eqnarray}
Our intention (tested below) is to assume that the vapor phase is
approximately ideal, $\beta\exmu_{\mathrm{vap}}\approx 0$. The result
Eq.~\eqref{eq:coex} applies to the coexisting liquid.   For the
thermodynamic states of  TABLE~\ref{t:II}, at the same temperature but
slightly different pressures, we apply the correction
\begin{eqnarray}
\beta \exmu_{\mathrm{liq}} \approx  -
\ln\left(\frac{\rho_{\mathrm{liq}}}{\rho_{\mathrm{vap}}}\right)
+\beta\left(\frac{\partial\mu}{\partial p}\right)_T\Delta p-
\ln\left(\frac{\rho}{\rho_{\mathrm{liq}}}\right)~.
\label{eq:corrected}
\end{eqnarray}
Of course $\beta\left(\partial\mu/\partial p\right)_T\Delta p = \beta
\Delta p/\rho_{\mathrm{liq}}$; at $T$=300K (TABLE~\ref{t:II}) this term
is about 0.003 and we neglect it.  The rightmost term of
Eq.~\eqref{eq:corrected} extracts the ideal contribution to the chemical
potential change. Collecting all, we find that
\begin{eqnarray}
-  \ln \lab\lab \chi \rab\rab_{0} \approx   -\ln\left(\frac{\rho}{\rho_{\mathrm{vap}}}\right) 
-  \beta\lab \varepsilon \rab 
      	- \beta^2\lab \delta \varepsilon^2 \rab/2 ~,
\label{eq:repul}
\end{eqnarray}
characterizes the net effect of intermolecular excluded volumes when the
vapor pressure is low.

Except for the highest temperature considered, the ideal estimate of the
vapor pressure is low and close to the experimental vapor pressure
(TABLE~\ref{tab:slabresults}), so the assumption of ideality of the
vapor is accurate then.  At $T$=600K, the vapor pressure is substantial
but the ideal estimate of the vapor pressure is still within about 20\%
of the experimental vapor pressure.

We therefore estimate the packing contribution
$-k_{\mathrm{B}}T\ln\lab\lab \chi \rab\rab_{0} \approx$ 15~kcal/mol at
300K, and 14~kcal/mol at 400K. Thus volume exclusion effects contribute
to the solvation free energies at the level of about 2/3rds of the
magnitude of the attractive interactions (TABLE~\ref{t:II}), of course
with opposite sign, when the vapor pressure is low. At 300K, $\lab\lab
\chi \rab\rab_{0} \approx 1\times 10^{-11}$.

Attractive interactions stabilize the liquid of course. van~der~Waals
attractions make a larger contribution to the mean binding energies
(TABLE~\ref{t:II}) than do electrostatic interactions. Contrariwise,
electrostatic contributions dominate van der Waals attractions in the
variances of binding energies.

If electrostatic interactions are considered solely, then at the lowest
temperature the mean and variance electrostatic contributions are
roughly in the 2:1 proportion that is a symptom of satisfactory Gaussian
models of solvation. If the distributions of binding energies are
precisely Gaussian, then
\begin{equation}
	\left\langle\left\langle \varepsilon  
 \right\rangle\right\rangle_0 = \left\langle \varepsilon 
 \right\rangle + \beta\left\langle \delta\varepsilon^2
 \right\rangle~.
\end{equation}
This follows from the general requirement that\cite{BPP}
\begin{eqnarray}
P\left(\varepsilon\right) = \me^{-\left(\varepsilon - \mu^{\mathrm{ex}}\right)}P^{\left(0\right)}\left(\varepsilon\right) 
\end{eqnarray}
where $P^{\left(0\right)}\left(\varepsilon\right)$ is the distribution
of binding energies for the uncoupled case associated with the 
$\left\langle\left\langle \ldots \right\rangle\right\rangle_0$
averaging.  Setting $\left\langle\left\langle \varepsilon
\right\rangle\right\rangle_0 = 0$ for electrostatic interactions, then
the 2:1 proportion of mean and variance contributions is clear in view
of the 1/2 in Eq.~\eqref{eqn:fnrg3}.  At the highest temperature
considered, the mean and variance contributions are roughly equal, so
Gaussian models of solvation are not supported then. The accuracy of the
inference of the packing contribution at $T$=600K is therefore less
convincing also.

\begin{table*}
	\begin{center} \caption{Bulk calculations at $p$ = 1~atm, $N$ = 600.
	 Energies are in kcal/mol.  $\kappa_T = \beta\left\langle \delta
	V^2\right\rangle/\left\langle V\right\rangle $. Experimental
	values\cite{Bottomley:1986wq,Marcus:1997td} for $\kappa_T$ at $T$ =
	25C are in the range 0.5-0.6 GPa$^{-1}$. Where not indicated
	explicitly, estimated statistical uncertainties are less than one
	(1) in the least significant digit given.  The thermal expansion
	coefficient implied by these results is 0.94$\times 10^{-3}$/K
	(experimental value:\cite{Piekarski:2010ks} 0.845$\times
	10^{-3}$/K).\label{t:II}}
\begin{tabular}{|c|c|c|c|c|c|}
\hline
	   $T$(K) 
	& $\rho$ (nm$^{-3}$) 
	&  $\left\langle\varepsilon\right\rangle$  
	& $\left\langle\varepsilon\right\rangle + \frac{\beta}{2}\left\langle\delta\varepsilon^2\right\rangle$
	& $ \kappa_T$ (GPa$^{-1}$)
	&  $\epsilon/\epsilon_0$ 

\\ \hline
270 & 7.5 & -33.2 &  -24.4  &  0.3 & 82.9 $\pm$ 2.9 
\\ \hline
300 & 7.3 & -32.0 &  -24.3 &  0.3  & 63.4 $\pm$ 2.2  
\\ \hline
320 & 7.2 & -31.2 &  -23.6 &  0.4 & 59.1 $\pm$ 2.3 
   \\ \hline

400 & 6.7 & -28.4 &  -21.1 &  0.5 &  41.8 $\pm$ 1.7 
     \\ \hline

600 & 5.5 & -22.0 &  -15.3  & 1.9 & 19.7 $\pm$ 1.3 
  \\ \hline
\end{tabular}
\end{center}
\end{table*}

\section{Conclusions}

Propylene carbonate (PC) does not spread completely on graphite, but at
ambient conditions it wets with a contact angle of 31$^\circ$. Molecular
dynamics simulations agree with this contact angle after 40\%  reduction of the
strength of graphite-C atom Lennard-Jones interactions  with the solvent, relative to
the models used initially.\cite{Yang:2009ty,Yang:2010hd} The simulation
of a nano-scale PC droplet on graphite displays a pronounced layering
tendency and an \emph{Aztex pyramid} structure for the droplet.
Extrapolation of the computed tensions of PC liquid-vapor interface
estimates the critical temperature of PC accurately to about 3\%.  PC
molecules lie flat on the PC liquid-vapor surface, and tend to project
the propyl carbon toward the vapor phase. Close PC neighbors stack
carbonate planes with the outer (O1) oxygen of one molecule snuggled
into the positively charged propyl end of another molecule so that
neighboring molecule dipole moments are approximately antiparallel.
The calculated thermal expansion coefficient and the dielectric constants for liquid PC 
agree well with experiment. The distribution of PC
molecule binding energies is closely Gaussian. Evaluation of the density
of the coexisting vapor then permits estimation of the excluded volume 
contribution to the PC chemical potential, and that contribution is
about 2/3rds of the magnitude of the contributions due to attractive
interactions, with opposite sign.

\section*{Acknowledgements} This work was supported by the National
Science Foundation under the NSF EPSCoR Cooperative Agreement No.
EPS-1003897, with additional support from the Louisiana Board of
Regents.

\section{APPENDIX: Methods}\label{Methods}

The GROMACS package\cite{Hess:2008tf} was used in all 
simulations. All the simulations were performed under periodic boundary
conditions, electrostatic interactions  calculated by the particle
mesh Ewald method with a grid spacing of 0.1 nm. Partial charges of PC
are those of Ref.~3.~
cutoff at 0.9~nm, and the temperature was maintained by the Nose-Hoover
thermostat.

For bulk PC systems, constant NPT conditions were adopted, treating 600
PC molecules initially positioned uniformly in a (4.4~nm)$^3$ cubic cell
utilizing Packmol.\cite{Martinez:2009di} The initial configuration was
energy-minimized, then simulations were carried out at $T$ = 270~K,
\ldots 600~K with $p$=1~atm, with 1~fs integration time step, and were
extended to 50~ns with first 10~ns discarded as aging.

Interfacial characteristics of liquid PC were investigated by molecular
dynamics of two-phase (liquid-vapor) systems in  slab geometry
(FIG.~\ref{fig:PurePC_3}). 600 PC molecules were positioned in a
$5.3\times5.3\times10.3$~nm$^3$ cell, $T$=300, 400, and 600K.  After
minimization, these systems were aged for 10~ns before a 10~ns
production equilibrium trajectory at 300~K. At 400~K and 600~K, 40~ns
equilibrium trajectories  were obtained. The interfacial tensions were
assessed (FIG.~\ref{fig:gamma04}) by differencing interfacial stresses
in the standard way, averaging through the production trajectories.

Configurations were sampled from each trajectory at every 0.005~ns for
further analysis.  The electrostatic contributions to the binding
energies  (TABLE~\ref{t:II}) were evaluated with the GROMACS reaction
field method.   Those electrostatic contributions were checked against
standard Ewald evaluations of electrostatic energies, and the
differences were typically about 0.1~kcal/mol. The differences of
generalized reaction field alternatives from the other methods were
substantially larger, \emph{i.e.} 1-2 kcal/mol.

Experimental applications almost always involve a racemic mixture of PC.
The calculations here were for the pure liquid of the enantiomer
FIG.~\ref{fig:pc_labelled}. This was because extensive initial
calculations made that choice,\cite{Yang:2009ty,Yang:2010hd} and no
results here are expected to be sensitive to that distinction.  
Explicit checking of a few cases in TABLE~\ref{tab:slabresults} and
TABLE~\ref{t:II} confirm that no results are changed significantly for
the racemic mixture.  Nevertheless, we expect subsequent results to
treat the racemic case.

\subsection{PC/graphite simulations}\label{ref:dispersion}

The PC/graphite simulations used 600 PC molecules and three layers of
graphite. Each layer has 9122 atoms, including 266 capping hydrogen
atoms at the edges, to make a square surface of about 15.5 nm by 15.5
nm. A liquid phase configuration of the PC molecules was placed near the
graphite surface and equilibrated for 5 ns. All simulations used
constant $TVN$ conditions, at $T$=300 K and a cubic (15.5~nm)$^3$
volume. To test how the PC/graphite interactions influence interfacial
properties, a series of simulations with different Lennard-Jones
$\varepsilon_{\mathrm{LJ}}$ values for the PC-graphite interactions were obtained, by
scaling $\varepsilon_{\mathrm{LJ}}$ by factors (0.25, 0.4, 0.5, 0.55, 0.6, 0.75 and
1). Interactions between all PC atoms and the graphite carbon atoms were all scaled.
Each system was simulated for 5 ns. For the optimal $\varepsilon_{\mathrm{LJ}}$
scaling, simulations were carried out for 10~ns. Simulations were
carried out  with and without constraints on the graphite surface
utilizing a harmonic restraint with force constant of 1000 kJ/(mol
nm$^2$). The simulations uses a cut-off value for equal to 0.9 nm.
Additional simulations were carried-out with a cut-off value of 1.2~nm.
The PC/graphite contact angle was found to insensitive to the cut-off
value or the use of restraints on the surface.  The indicated 40\%
reduction in the strength of graphite-C Lennard-Jones interactions
may be the simplest adjustment that brings the simulation results
into consistency with the contact angle observation.

\subsection{Droplet silhouette}
The  droplet silhouette (FIG.~\ref{fig:aztexPyramid}) was obtained
in the following way:\cite{Werder:2003ve} On the basis of the simulation
data, the mass density was binned. A two dimensional grid was
used in cylindrical coordinates with $z$ perpendicular to the graphite
plane and $r$ the radial coordinate.  The $z$  axis contained the
centroid of the droplet. The function
\begin{equation}
\rho(r,z) = \frac{\rho_{\mathrm{liq}}}{2} \left\{1- 
\tanh\left\lbrack\frac{r-r_0(z)}{d(z)}\right\rbrack\right\}~,
\end{equation}
which acknowledges that the $\rho_{\mathrm{vap}}\approx 0$, was fit
to the binned mass density. Thus the width of the interfacial profile,
$d(z)$, and the position of the interface, $r_0(z)$ were obtained for
each $z$ layer. Since the density exhibits distinct layering parallel to
$z$ (FIG.~\ref{fig:layering}), this procedure is particularly effective where
analysis of the density in constant-$z$ slices is natural, for $0 < z <
1.9$~nm.

For $z > 1.9~\mathrm{nm}$, where the $0<r<2~\mathrm{nm}$ portion of the
fluid interface is roughly parallel to the $z$=constant (graphite)
surface, constant-$r$ slices of the density were similarly fit to the  
function
\begin{equation}
\rho(r,z) = \frac{\rho_{\mathrm{liq}}}{2} \left\{1- 
\tanh\left\lbrack\frac{z-z_0(r)}{d(r)}\right\rbrack\right\}~.
\end{equation}
In view of FIG.~\ref{fig:layering}, this describes the outer
(largest $r$) behavior, where the density is decreasing through 
$\rho_{\mathrm{liq}}/2$. The two approaches give the same values near a
crossing point $r\approx 2$~nm. Combing these two fits gives
FIG.~\ref{fig:aztexPyramid}.

\subsection{Windowed sampling for calculations of coexisting vapor densities}\label{WHAM}

The coexisting vapor densities were evaluated by stratification on the
basis of the Weighted Histogram Analysis Method
(WHAM)\cite{Kumar:1992p14782,AlanGrossfield} to concentrate sampling on
the low densities of the vapor phase.  Windowed calculations were
performed with 600 PC molecules at temperatures 300K and 400K adopting
the methods described above. Initial configuration for each window was
obtained by pulling a PC molecule from the center of mass of the slab to
deep into vapor phase at constant rate of 0.01~nm/ps. In the 300K case,
15 windows, spaced by 0.1~nm in bulk liquid phase and 0.05~nm near
interface and vapor phase, were treated. Trajectories of 5 ns/window
(after 1 ns of minimization and 5~ns of aging) were used to reconstruct
density profiles. For the 400K calculation, 41 windows of 0.1~nm uniform
spacing were used.

The harmonic potential energy ${U = k(z-z_0)^2}$ windowing function was
employed, with  $z$ is the instantaneous distance of the center of mass
of the pulled PC molecule from center of mass of the slab and $z_0$, the
designated minimum of $U$, identifying the window position. $k$ = 4000
$\mathrm{kJ/mol/{nm}^2}$ (for 300K) and $k$=6000
$\mathrm{kJ/mol/{nm}^2}$ (for 400 K).

\subsection{Liquid PC dielectric constant}

The dielectric constant of uniform PC liquid is evaluated following
standard simulation methods.\cite{Yang:2010hd,Wu:2006bg} The calculation
treated 600 PC molecules under periodic boundary conditions. Resulting
values were averaged from the 40~ns production trajectories at constant
pressure of 1~atm.

\newpage
%

\end{document}